\documentclass[a4paper,11pt]{article}
\pdfoutput=1 

\usepackage{jcappub} 
                     
\usepackage{txfonts} 

\newcommand{\bra}{\langle}
\newcommand{\ket}{\rangle}
\newcommand{\tr}{\mathrm{tr}}

\title{\boldmath Environmental dependence of ellipticity correlation functions of intrinsic alignments}

\author[a,b,c]{Robert Reischke\note{Corresponding author.}}
\author[a]{and Bj\"orn Malte Sch\"afer}

\affiliation[a]{Astronomisches Rechen-Institut, Zentrum f{\"u}r Astronomie der Universit{\"a}t Heidelberg, Philosophenweg 12, 69120 Heidelberg, Germany}
\affiliation[b]{Institut f\"ur Kernphysik, Karlsruher Institut f\"ur Technologie, 76344 Eggenstein-Leopoldshafen, Germany}
\affiliation[c]{Physics Department, Technion, 3200003 Haifa, Israel}

\emailAdd{r.reischke@technion.ac.il}

\abstract{In this work we investigate the environmental dependence of the intrinsic ellipticity correlations in cosmic shear surveys. We use the quadratic and linear alignment model to describe the contributions by spiral and elliptical galaxies, respectively. The density field is in both cases described by a Gaussian random field and ellipticity correlation functions that are conditional on the environment of the galaxies are constructed by sampling random values for the tidal tensor and inertial tensor. The covariance of the Gaussian random process from which the tensor entries are drawn is decomposed by means of a spherical Fourier-Bessel transformation of the density field. The dependence on environment is modelled by the number of positive eigenvalues of the tidal tensor, which allows a differentiation between voids, sheets, filaments and superclusters. We find that elliptical galaxies align strongest in elongated structures such as sheets and filaments with an amplitude almost an order of magnitude higher compared to the alignment in clusters or voids. In contrast to this, spiral galaxies align equally strong in all environments. Cross-alignments between different environments are smaller than the respective auto-correlations subject to the Cauchy-Schwarz inequality which is an effective bound on their amplitude. Furthermore, we find misalignment between inertial and tidal tensor to be stronger in anisotropic regions compared to clusters or voids.
While the imprint of weak lensing on galaxy ellipticities is agnostic about the environment, using environment information can help to distinguish between lensing and the intrinsic alignment signal. }

\begin{document}
\maketitle
\flushbottom

\section{Introduction}\label{sec:intro}
Intrinsic alignments are one of the major systematic effects in future cosmological surveys such as Euclid or the Large Synoptic Sky survey LSST. They introduce physical shape correlations of galaxies and therefore mimic a weak gravitational lensing signal \citep[see e.g.][]{bartelmann_weak_2001, refregier_weak_2003, kilbinger_cosmology_2015}. The intrinsic alignment signal is expected to make up just under 10 percent of the total lensing signal of Euclid over a wide range of multipoles, assuming standard alignment models for both major galaxy types with a conservative choice of alignment model parameters.

The intrinsic shape correlations arise due to the gravitational interaction of the galaxies with the ambient large-scale structure \citep[][for reviews]{jing_intrinsic_2002, heavens_intrinsic_2000, croft_weak-lensing_2000, catelan_correlations_2001, mackey_theoretical_2002, lee_intrinsic_2005, troxel_intrinsic_2015, joachimi_galaxy_2015, kiessling_galaxy_2015}. Intrinsic alignments have been studied extensively both with analytical methods \cite[see][]{tugendhat_angular_2017, blazek_beyond_2017, merkel_parameter_2017, merkel_intrinsic_2013} and simulations \citep[e.g.][]{codis_intrinsic_2015, hilbert_intrinsic_2016, wei_euclid_2018}. There are by now robust detections of intrinsic shape correlations of elliptical galaxies, including a cross-correlation between intrinsic shapes of elliptical galaxies and weak lensing, but todate, detections of alignments of spiral galaxies are still inconclusive.

Theoretical modelling of intrinsic alignments assumes different alignment mechanisms for the two major galaxy types. Elliptical galaxies receive their intrinsic shape due to tidal forces exerted on the virialised halo of stars and dark matter particles. Due to proportionality between the galaxy shape and the tidal field the latter is called the linear alignment model \citep{hirata_galaxy-galaxy_2004, blazek_testing_2011, blazek_tidal_2015, camelio_origin_2015}. Spiral galaxies are modelled differently in the sense that their apparent shape is dictated by the inclination angle under which the galactic disc is viewed. Therefore, the decisive quantity is their angular momentum \citep{lee_cosmic_2000, lee_galaxy_2001}, which is thought to coincide with the symmetry axis of the galactic disc. The tidal torquing model relates the angular momentum of the galaxy's host halo to a misalignment between the inertia tensor and tidal shear tensor \citep{bailin_internal_2005}. Since this model depends quadratically on the density contrast it is referred to as quadratic model \citep{schaefer_review:_2009, schaefer_galactic_2012}.

There are many physical and statistical differences between the alignments of elliptical galaxies, of spiral galaxies and of the gravitational lensing effect which can be used to enhance or suppress signals. Alignments of elliptical galaxies dominate at multipole while spiral galaxies only contribute at high multipoles. Alignments are local, implying that they would not show up in cross correlations between different tomography bins, in contrast to weak lensing. Spiral galaxies would exhibit non-Gaussian statistics in their ellipticity field even when generated from a purely Gaussian random field, while both lensing and the alignment of elliptical galaxies provide a linear mapping of the underlying density field irrespective of its statistical properties. While elliptical galaxies and lensing would not be able at lowest order to generate $B$-modes in the ellipticity field, this is markedly different for the case of spiral galaxies. Apart from statistical and tomographic differences \citep[as listed by][]{tugendhat_angular_2017}, the different redshift weighting of lensing and intrinsic alignments suggests a possible mitigation strategy by employing redshift weightings, which has been investigated in nulling and boosting-schemes.

Simulations have shown that massive halos hosting elliptical galaxies align along filaments and sheets \citep{patiri_alignment_2006, brunino_orientation_2007}. Furthermore, galaxy-halo alignments are tested with hydrodynamical simulations \citep{chisari_galaxy-halo_2017, xia_halo_2017}. The alignment of galaxies with filaments has be subject of a number of studies \citep{lee_cosmic_2000, catelan_correlations_2001, codis_intrinsic_2015}. Most recently, \citep{codis_galaxy_2018} investigated the alignment of dark matter halo shapes with walls in the LSS and a corresponding perpendicular alignment of halo angular momenta. These studies motivate us to pursue the question whether shape correlations of galaxies depend on the environment, because in both alignment models, the random process generating tidal shears is restricted by the large-scale dynamics of the structure in which the galaxy resides.

Therefore, in this work we investigate how the intrinsic ellipticity correlations of both spiral and elliptical galaxies depend on the environment the selected pairs are exposed to. In fact, the central object of our studies are conditional ellipticity correlation functions, where the ellipticities are taken from a subset of environments, such that in a marginalisation over all environments one would recover a straightforward measurement of shape correlation. While intrinsic shape correlations should depend on the environment, the gravitational lensing effect, which provides an integrated measure of tidal shear fields, can not depend on the galaxy environment and this difference can potentially open a window for differentiating between alignment mechanisms and the gravitational lensing on the basis of environmental dependence. 

We use the linear and quadratic alignment model to describe shape correlations. The correlations of the gravitational tidal field together with the mass tensor are calculated from the fluctuation statistics of a Gaussian random field in a sampling process \citep{schaefer_galactic_2012, schaefer_angular_2015}. Here, we rely on linear perturbation theory and are therefore able to sample from the joint distribution of tidal fields and the components of the mass tensor at two different points, on which we can impose criteria that characterise the environment. Those criteria are taken to be the number of positive eigenvalues of the tidal shear tensor, which would be indicative of the location of the halo inside a void, sheet, filament or cluster. In addition, we will show results on cross correlations of ellipticities of galaxies that reside in different environments.

The structure of the paper is as follows: In \autoref{sec:shapecorrelations} we review the linear and quadratic alignment model as well as well as the correlations of tidal fields the density and its derivatives in a spherical Fourier-Bessel basis. \autoref{sec:sampling} describes the details of the sampling process, in which we determine the galaxy ellipticities from the fluctuation statistics of a Gaussian random field. We present our results on auto- and cross correlation functions in \autoref{sec:results} and conclude in \autoref{sec:summary}. Our reference cosmology is a $\Lambda$CDM-model with the parameter choice $\Omega_m$, $\sigma_8$, $h$, $n_s$ and $w=-1$, and we work under the assumption of linearly evolving structures with Gaussian statistics. While the restriction on a comparatively simple cosmological model is not serious as intrinsic alignments are rather insensitive to the fundamental cosmological model, in particular to the dark energy-section, Gaussian statistics is only valid for linearly evolving structures.

\section{Shape correlations}\label{sec:shapecorrelations}
Intrinsic shape correlations between galaxies are a consequence of their formation process and of gravitational interaction of the galaxies with the ambient large-scale structure. The interaction type depends on the galaxy under consideration: Elliptical galaxies will orient themselves in a preferred direction defined by the local gravitational tidal field, and their the response in their shapes to the tidal shear is to a good approximation linear. On the other hand shape correlations of spiral galaxies are due to correlated orientations of the galactic disk, which in turn are induced by angular momentum correlations. In this sense, the intrinsic shape correlation of galaxies inherit the statistics of the underlying cosmic density field in both models. However, the alignment will in general depend on the environment of the considered galaxy pair, because imposing a requirement on e.g. the number of positive eigenvalues of the tidal shear tensor, changes the properties of the random process that generates values for the tidal field and in the case of spiral galaxies, of the inertia tensor as well. We consider the number of positive eigenvalues of the tidal shear tensor to be a very straightforward discriminant between voids (all eigenvalues of $\partial^2\Phi$ are negative), sheets (two negative and one positive eigenvalue), filaments (one negative and two positive eigenvalues) and superstructures (all positive eigenvaleus), as they reflect the large-scale dynamics of the local environment.
 
In this section we will describe the statistics of the intrinsic shape correlations derived from statistics of the large-scale structure, which particularly tests halo-alignment models \citep{schneider_halo_2010, joachimi_intrinsic_2013, joachimi_intrinsic_2013-1}. This is done by setting up an environment-dependent Gaussian random process for drawing tidal shear values, and computing their variance if a restriction is in place.

\subsection{Statistics of the density field}
The statistics of the cosmic density field, $\delta(\boldsymbol{x})$, are described by its correlation function or the power spectrum, $P(k)$:
\begin{equation}\label{eq:power_spectrum_definition}
\langle\delta(\boldsymbol{k})\delta^*(\boldsymbol{k^\prime})\rangle = (2\pi)^3\delta_\mathrm{D}^3(\boldsymbol{k} -\boldsymbol{k^\prime})P(k)\; ,
\end{equation}
where $\delta_\mathrm{D}$ is the Dirac distribution and we used the Fourier convention
\begin{equation}\label{eq:fourier_convention}
\delta(\boldsymbol{k}) = \int\mathrm{d}^3 x\; \delta(\boldsymbol{x}) \exp(-\mathrm{i}\boldsymbol{kx})\; 
\end{equation}
For the power spectrum we assume the form
\begin{equation}\label{eq:cdm_power}
P(k,a) \propto D_+^2T^2(k,a) k^{n_\mathrm{s}}\;,
\end{equation}
with the linear growth factor $D_+$ given as a solution to the following differential equation
\begin{equation}\label{eq:growth_factor}
\frac{\mathrm{d}^2D_+}{\mathrm{d}a^2} + \frac{\mathrm{d}D_+}{a\mathrm{d}a}\left(3+ \frac{\mathrm{d}\log H}{\mathrm{d}\log a}\right) = \frac{3}{2a^2}\Omega_\mathrm{m}(a)D_+\;.
\end{equation}
With the Hubble function, $H(a)$ and the matter density $\Omega_\mathrm{m}(a)$ at scale factor $a$. The transfer function, $T(k,a)$ is taken from \citep{bardeen_statistics_1986} and the normalization is by measuring the fluctuation amplitude at a scale of $R=8 \;\mathrm{Mpc}/h$ today, $D_+(a=1)=1$:
\begin{equation}\label{eq:variance_R}
\sigma_R^2 = \frac{1}{2\pi^2}\int \mathrm{d}k\; k^2 W^2(kR) P(k),
\end{equation} 
with a top-hat filter function, $W(kR)$ in real space. From there, the spectral moments for the mass scale $M$, can be calculated as
\begin{equation}\label{eq:spectral_moments}
\sigma^M_j = \frac{1}{2\pi^2}\int \mathrm{d}k \; k^{2j+2} W_M^2(k) P(k),
\end{equation}
where $W_M(k)$ is a Gaussian low-pass filter with length scale $L = [4/(3M)\pi\rho_0(a)]^{-1/3}$, where $\rho_0(a) = \Omega_\mathrm{m}(a)\rho_\mathrm{crit}(a)$ is the mean cosmic density. We use a Gaussian filter rather than a top-hat filter in order to avoid convergence issues for high values of $j$.

\subsection{Alignments in tidal fields in Lagrangian perturbation theory}
The growth of perturbation can also be treated using Lagrangian perturbation theory \citep{zeldovich_gravitational_1970, white_angular_1984, catelan_evolution_1996-1, crittenden_spin-induced_2001}. In the Zel'dovich approximation the trajectory of a particle in co-moving coordinates is given by
\begin{equation}\label{eq:Zeld}
 \boldsymbol{x} =  \boldsymbol{q}-D_+(t) \boldsymbol{\nabla} \psi \ \ \Rightarrow \ \ \boldsymbol{\dot q} = -\dot D_+(t) \boldsymbol{\nabla}\psi\;.
\end{equation}
Here ${\psi}$ is the displacement field, which can be related to the density contrast by virtue of the Poisson equation, $\Delta \psi = \delta$. Using this relation the tidal fields of the LSS and the curvature of the density field can be related to the Fourier modes $\delta(\mathrm{k})$ whose statistics is known:
\begin{align}
\label{eq:delta}
\delta (\boldsymbol{x}) & = \int\frac{\mathrm{d}^3 k}{(2\pi)^3}\delta(\boldsymbol{k}) \exp(\mathrm{i}\boldsymbol{kx})\; , \\
\label{eq:delta_x}
\partial_\alpha \delta (\boldsymbol{x}) & = \mathrm{i}\int\frac{\mathrm{d}^3 k}{(2\pi)^3}k_\alpha\delta(\boldsymbol{k}) \exp(\mathrm{i}\boldsymbol{kx})\; , \\
\label{eq:delta_xx}
\delta_{\alpha\beta}\equiv\partial_\alpha\partial_\beta \delta (\boldsymbol{x}) & = -\int\frac{\mathrm{d}^3 k}{(2\pi)^3}k_\alpha k_\beta\delta(\boldsymbol{k}) \exp(\mathrm{i}\boldsymbol{kx})\; , \\
\label{eq:psi_xx}
\psi_{\alpha\beta} \equiv \partial_\alpha\partial_\beta \psi (\boldsymbol{x}) & = -\int\frac{\mathrm{d}^3 k}{(2\pi)^3}\frac{k_\alpha k_\beta}{k^2}\delta(\boldsymbol{k}) \exp(\mathrm{i}\boldsymbol{kx})\; .
\end{align}
The high powers of $k$ shows that the correlation length in $\delta_{\alpha\beta}$ is much shorter than the correlation length in the tidal field $\psi_{\alpha\beta}$. It should also be noted that the density field is degenerate with the trace of the tidal tensor $\psi_{\alpha\beta}$, hence $\delta$ is not an independent quantity but a derived one.

The tensor of inertia $\boldsymbol{I}$ can be related to $\delta_{\alpha\beta}$ at a particular point $\boldsymbol{x}_0$ in the eigenframe of the mass tensor which is given by $-\delta_{\alpha\beta}$. At its peak we can describe the density profile by a Taylor expansion in the following way \citep{catelan_evolution_1996}
\begin{equation}\label{eq:density_at_peak}
\delta(\boldsymbol{x}) = 
\delta(\boldsymbol{x}_0) - \frac{1}{2}\sum_i^3 \lambda_i (\boldsymbol{x}-\boldsymbol{x}_0)_i^2\;,
\end{equation}
with the eigenvalues $\lambda_i$ of the mass tensor. The boundary, $\partial\Gamma$, of the peak region $\Gamma$ can be defined as an iso-density contour, while its height can be expressed in units of the smoothed standard deviation, $\sigma^M_0$, such that $\delta(\boldsymbol{x_0}) = \nu\sigma^M_0$. The tensor of inertia is the mass-tensor's eigenframe is then given by
\begin{equation}\label{eq:inertial_tensor}
I_{\alpha\beta} = \frac{\eta_0}{5}\Gamma \mathrm{diag}\left(A_y^2+A_z^2, A_x^2 + A_z^2, A_y^2+A_x^2 \right)\; ,
\end{equation}
with $\eta_0 = \rho a^3$ and the volume of $\Gamma$:
\begin{equation}
\Gamma = \frac{4\pi}{3}A_xA_yA_z\;,
\end{equation}
where $A_\alpha$ describe the semi-axes of the ellipsoid:
\begin{equation}
A_\alpha = \sqrt{\frac{2\nu\sigma_0}{\lambda_\alpha}}\; .
\end{equation}

\subsection{Tidal fields in spherical coordinates}
As in \citep{reghos_density_1995} and \citep{heavens_correlation_1999} correlations between the density and its derivatives are expressed in a spherical basis. Considering two points situated symmetric around the origin on the $z$-axis and separated by a distance $r$ the correlations take a simple form in the complex basis $y_{\ell m}^n)(\boldsymbol{x})$:
\begin{equation}\label{eq:spherical_basis}
y^n_{\ell m}(\boldsymbol{x}) = (4\pi)^{\frac{1}{2}} \frac{\mathrm{i}^{\ell +2n}}{\sigma_{\ell+2n}}\int\frac{\mathrm{d}^3k}{(2\pi)^3}k ^{\ell +2n} \delta(\boldsymbol{k})Y_{\ell m}(\boldsymbol{\hat{k}}) \exp(\mathrm{i}\boldsymbol{k}\boldsymbol{x})\; .
\end{equation}
The $y^n_{\ell m}$ frame is related to the physical frame in the following way:
\begin{equation}\label{eq:trafo_y_to_phys}
\begin{split}
\sigma_2 y^1_{00}(\boldsymbol{x}) &= \left( \delta_{xx}(\boldsymbol{x}) + \delta_{yy}(\boldsymbol{x}) + \delta_{zz}(\boldsymbol{x})\right)\;, \\
\sigma_0 y^0_{00}(\boldsymbol{x}) & =\delta(\boldsymbol{x}) = \Delta\psi(\boldsymbol{x})\;, \\
\sigma_1 y^0_{10}(\boldsymbol{x}) & =\sqrt{3}\delta_z(\boldsymbol{x})\; ,\\
\sigma_2 y^0_{20}(\boldsymbol{x}) &= -\sqrt{5/4} \left( \delta_{xx}(\boldsymbol{x}) + \delta_{yy}(\boldsymbol{x}) - 2 \delta_{zz}(\boldsymbol{x})\right)\;, \\
\sigma_0 y^{-1}_{20}(\boldsymbol{x}) &= -\sqrt{5/4}\left( \psi_{xx}(\boldsymbol{x}) + \psi_{yy}(\boldsymbol{x}) -2 \psi_{zz}(\boldsymbol{x})\right)\;, \\
\sigma_1 y^0_{11}(\boldsymbol{x}) & = -\sqrt{3/2} \left(\delta_x(\boldsymbol{x}) + \mathrm{i}\delta_y(\boldsymbol{x})\right)\; , \\
\sigma_2 y^0_{21}(\boldsymbol{x}) &= -\sqrt{15/2} \left( \delta_{xz}(\boldsymbol{x}) + \mathrm{i}\delta_{yz}(\boldsymbol{x}) \right)\;, \\
\sigma_0 y^{-1}_{21}(\boldsymbol{x}) &= -\sqrt{15/2} \left( \psi_{xz}(\boldsymbol{x}) + \mathrm{i}\psi_{yz}(\boldsymbol{x}) \right)\;, \\
\sigma_1 y^0_{1-1}(\boldsymbol{x}) & = \sqrt{3/2} \left(\delta_x(\boldsymbol{x}) - \mathrm{i}\delta_y(\boldsymbol{x})\right)\; , \\
\sigma_2 y^0_{2-1}(\boldsymbol{x}) &= \sqrt{15/2} \left( \delta_{xz}(\boldsymbol{x}) - \mathrm{i}\delta_{yz}(\boldsymbol{x}) \right)\;, \\
\sigma_0 y^{-1}_{2-1}(\boldsymbol{x}) &= \sqrt{15/2} \left( \psi_{xz}(\boldsymbol{x}) - \mathrm{i}\psi_{yz}(\boldsymbol{x}) \right)\;, \\
\sigma_2 y^0_{22}(\boldsymbol{x}) &= \sqrt{15/8} \left( \delta_{xx}(\boldsymbol{x}) - \delta_{yy}(\boldsymbol{x}) + 2 \mathrm{i}\delta_{xy}(\boldsymbol{x})\right)\;, \\
\sigma_0 y^{-1}_{22}(\boldsymbol{x}) &= \sqrt{15/8} \left( \psi_{xx}(\boldsymbol{x}) - \psi_{yy}(\boldsymbol{x}) + 2\mathrm{i}\psi_{xy}(\boldsymbol{x}) \right)\;, \\
\sigma_2 y^0_{2-2}(\boldsymbol{x}) &= \sqrt{15/8} \left( \delta_{xx}(\boldsymbol{x}) - \delta_{yy}(\boldsymbol{x}) - 2 \mathrm{i}\delta_{xy}(\boldsymbol{x})\right)\;, \\
\sigma_0 y^{-1}_{2-2}(\boldsymbol{x}) &= \sqrt{15/8} \left( \psi_{xx}(\boldsymbol{x}) - \psi_{yy}(\boldsymbol{x}) - 2\mathrm{i}\psi_{xy}(\boldsymbol{x}) \right)\;.
\end{split}
\end{equation}
Here the symmetry of the $y^n_{\ell m}$ basis 
\begin{equation}\label{eq:symmetry_spherical}
y^{n}_{\ell m}(\boldsymbol{x})^* = (-1)^m y^n_{\ell, -m}(\boldsymbol{x})\;,
\end{equation}
was exploited. The symmetry is inherited from the reality of the density field and is crucial when inverting the system of equations (\ref{eq:trafo_y_to_phys}) to close the system. We now bundle the values of $y^n_{\ell m}$ as in Eq (\ref{eq:trafo_y_to_phys}) into a vector $\boldsymbol{y}_\mathrm{sph}$. Likewise we write the physical quantities into another vector $\boldsymbol{y}_\mathrm{phys}$:
\begin{equation}
\boldsymbol{y}^\mathrm{T}_\mathrm{phys} \equiv \left(\delta_x,\delta_y,\delta_z,\delta_{xx},\delta_{yy},\delta_{zz},\delta_{xy},\delta_{xz},\delta_{yz},\psi_{xx},\psi_{yy},\psi_{zz},\psi_{xy},\psi_{xz},\psi_{yz}\right)\; .
\end{equation}

\subsection{Correlations in the spherical basis}
We give the correlations in the $y^n_{\ell m}$ basis as calculated in \citep{reghos_density_1995}. The covariance matrix can be written in the following simplified block-matrix form,
\begin{equation}\label{eq:correlation_matrix}
\boldsymbol{Y}  =
\begin{pmatrix}
\boldsymbol{A} & \boldsymbol{C} \\
\boldsymbol{C}^+ & \boldsymbol{A} \\
\end{pmatrix}\; ,
\end{equation}
where $\boldsymbol{A}$ is the auto-correlation matrix and $\boldsymbol{C}$ is the cross-correlation matrix with components
\begin{equation}\label{eq:correlation_matrix_components}
A_{ij} = \langle y_i(\boldsymbol{x}) y_j^*(\boldsymbol{x})\rangle\; ,\quad C_{ij} = \langle y_i (\boldsymbol{x}) y^*_j(\boldsymbol{x}^\prime)\rangle\;,
\end{equation}
where we collected the indices $(n,\ell,m)$ to a new index $i$. Therefore, both $\boldsymbol{A}$ and $\boldsymbol{C}$ make up a $15\times 15$-matrix, respectively. In contrast to $\boldsymbol{A}$, $\boldsymbol{C}$ is a complex matrix, and the cross denotes complex conjugation followed by matrix transposition.

By exploiting the orthogonality relation of the spherical harmonics the auto-correlation matrix assumes a very simple form:
\begin{equation}\label{eq:auto_correlation}
A^{nn'}_{mm'\ell\ell'} = A^{nn'}_{\ell m} \delta_{\ell \ell'}\delta_{m m'} =  (-1)^{n-n'} \frac{\sigma^2_{\ell +n +n'}}{\sigma_{\ell +2n}\sigma_{\ell+2n'}}\delta_{\ell\ell'}\delta_{mm'}\;, 
\end{equation}
with the spectral moments defined in eq. (\ref{eq:spectral_moments}). 
Successive application of Rayleigh's formula:
\begin{equation}\label{eq:rayleigh}
\exp(\mathrm{i}\boldsymbol{kr}) = 4\pi \sum_{L= 0}^\infty \mathrm{i}^Lj_L(kr) \sum_{M=-L}^LY_{LM}(\hat{\boldsymbol{r}})^*Y_{LM}(\hat{\boldsymbol{k}})\; ,
\end{equation}
and inserting the definition of the $y^n_{\ell m}$ coefficients, the cross-correlation components can be brought into the following form:
\begin{equation}\label{eq:cross_correlation}
\begin{split}
C^{nn'}_{\ell\ell'mm'}(r)  = & \ \delta_{mm'}\frac{(-1)^{m+n-n'}}{\sigma_{\ell +2n}\sigma_{\ell'+2n'}}\sum_{L = |\ell-\ell'|}^{\ell+\ell'} (2L+1)\mathrm{i}^{L+\ell-\ell'}K_{L, \ell+\ell'+2(n+n'+1)}(r)\\ & \times \sqrt{(2\ell+1)(2\ell'+1)}
\begin{pmatrix}
\ell & \ell' & L \\ 0 & 0 & 0
\end{pmatrix}
\begin{pmatrix}
\ell & \ell' & L \\ m & -m & 0
\end{pmatrix}\;.
\end{split}
\end{equation}
Here we used that
\begin{equation}
\int\mathrm{d}\Omega \;Y_{\ell_1m_1}(\hat{\boldsymbol{k}})Y_{\ell_2m_2}(\hat{\boldsymbol{k}})Y_{\ell_3m_3}(\hat{\boldsymbol{k}}) = (-1)^{m_2} \sqrt{\frac{\prod_{i=1}^3(2\ell_i+1)}{4\pi}} \begin{pmatrix}
\ell_1 & \ell_2 & \ell_3 \\ 0 & 0 & 0
\end{pmatrix}
\begin{pmatrix}
\ell_1 & \ell_2 & \ell_3 \\ m_1 & -m_2 & m_3
\end{pmatrix}\;,
\end{equation}
where the Wigner-$3j$ symbol was introduced. Furthermore, the spectral moments weighted with spherical Bessel functions are abbreviated as
\begin{equation}
K_{\ell m} (r) = \frac{1}{2\pi^2}\int \mathrm{d}k\; k^mj_\ell (kr)P(k)\; .
\end{equation}
We show the correlation function (\ref{eq:cross_correlation}) on the left side of \autoref{fig:ellipticals_contribution} for different combinations of $n,\ell,m$.

\subsection{Ellipticity correlations}
The intrinsic shapes of galaxies can be described by the tidal shearing-model \citep{tugendhat_angular_2017, blazek_tidal_2015}, and the tidal-torquing model \citep{catelan_evolution_1996, theuns_angular_1997, catelan_correlations_2001, croft_weak-lensing_2000,crittenden_spin-induced_2001,crittenden_discriminating_2002} for elliptical and spiral galaxies, respectively.

In the tidal shearing-model an elliptical galaxy is virialised system which consists out of stars which are characterized by a constant velocity dispersion. Introducing an anisotropy parameter into the Jeans equation allows to investigate the influence of tidal field on the virialised structure by Taylor expanding the density profile as
\begin{equation}\label{eq:density_profile_elliptical}
\rho\propto \exp\left(-\frac{\psi(r)}{\sigma^2}\right)\left(1-\frac{1}{2\sigma^2}\psi_{\alpha\beta}x^\alpha x^\beta \right)\; .
\end{equation}
The proportionality constant, $D$, does not depend on mass or redshift \citep{tugendhat_angular_2017,piras_mass_2017}. This deformation of the iso-density contours of a collection of stars in virial equilibrium gives rise to an intrinsic shape which inherits its statistical properties from the statistics of the tidal shear field. The ellipticity will only depend on the orientation, i.e. the eigen system, and the magnitude of the external shear field. In a coordinate system where the ellipse falls into the $x,y$-plane, i.e. where we defined the $z$-axis to be the line-of-sight, the complex ellipticity $\epsilon$ can be expressed as
\begin{equation}\label{eq:linear_model}
\epsilon = \epsilon_+ + \mathrm{i}\epsilon_\times = D(\psi_{xx} -\psi_{yy} + 2\mathrm{i} \psi_{xy})\; .
\end{equation}
Since the ellipticity is linear in the tidal field and thus also in the density field, the tidal shearing-model is also dubbed linear model.

In contrast, the intrinsic shape correlations of spiral galaxies can be described by the correlation of their angular momenta \citep{croft_weak-lensing_2000,crittenden_spin-induced_2001}. The angular momentum of spiral galaxies are correlated due to correlated initial conditions in this picture.
Consequently, observed ellipticities are also correlated due to the inclination of the galactic disk. The complex ellipticity in this case can be brought into the following form:
\begin{equation}\label{eq:quadratic_model}
\epsilon = \frac{\hat{L}_x^2-\hat{L}_y^2}{1+ \hat{L}_z^2} + 2\mathrm{i}\frac{\hat{L}_x\hat{L}_y}{1+\hat{L}_z^2}\; ,
\end{equation}
where $\boldsymbol{\hat{L}}$ denotes the normalized angular momentum whose components can be obtained in the tidal torquing model \citep{doroshkevich_space_1970,white_angular_1984,catelan_evolution_1996,crittenden_spin-induced_2001}  via
\begin{equation}\label{eq:ang_moment}
L_\alpha =a^3H(a)\frac{\mathrm{d}D_+}{\mathrm{d}a}\sum_{\beta,\gamma,\sigma}\epsilon_{\alpha\beta\gamma} I_{\beta\sigma}\psi_{\sigma\gamma}\;.
\end{equation}
Clearly, angular momentum is only generated if the tensor of inertia does not align with the tidal tensor. If they shared a common eigensystem, the product $\sum_\sigma I_{\beta\sigma}\psi_{\sigma\gamma}$ is symmetric and the contraction with the Levi-Civit{\`a}-symbol yields zero.

Finally it should be noted that the observed shape correlation functions are related to (\ref{eq:linear_model}) and (\ref{eq:quadratic_model}) by line of sight integrals with a suitable weight function. The combined ellipticity of a galaxy image is then given, at lowest order, by $\epsilon = g+ \epsilon_\mathrm{I}$, where $g$ labels the contribution from gravitational lensing and $\epsilon_\mathrm{I}$ is the intrinsic shape contribution. $\epsilon_\mathrm{I}$ itself splits up into a part intrinsic to the galaxy and another part due to its interaction with the large scale structure which this work is about.

\section{Sampling process}\label{sec:sampling}
Let us now assume that we are given two points in the density field, labelled $\boldsymbol{x}$ and $\boldsymbol{x}'$ respectively. Random numbers, $\boldsymbol{y}_\mathrm{sph}$, can be generated at the two points in the $y^n_{\ell m}$ basis by drawing random number from a Gaussian distribution with covariance $\boldsymbol{Y}$ from Eq. (\ref{eq:correlation_matrix}):
\begin{equation}\label{eq:Gaussian}
\mathcal{N}(\boldsymbol{0},\boldsymbol{Y})\quad \to\quad \boldsymbol{y}_\mathrm{sph}\;.
\end{equation}
Inverting the system of equations (\ref{eq:trafo_y_to_phys}), $\boldsymbol{y}_\mathrm{sph}$ can be mapped onto the physical quantities, i.e. the density field its derivatives and the tidal field, $\boldsymbol{y}_\mathrm{phys}$. Now the mass tensor and the tidal tensor can be calculated at $\boldsymbol{x}$ and $\boldsymbol{x}'$. 
The eigenvalues of $\boldsymbol{\psi}$, $\lambda_{\psi,i}$ will indicate the environment the object is situated at:
\begin{align*}\label{eq:environment}
\text{void} & \quad\leftrightarrow\quad 3 \ \text{of the } \lambda_{\psi,i} \ \text{are positive}\;, \\
\text{sheet} & \quad\leftrightarrow\quad 2 \ \text{of the } \lambda_{\psi,i} \ \text{are positive}\;, \\
\text{filament} & \quad\leftrightarrow\quad 1 \ \text{of the } \lambda_{\psi,i} \ \text{is positive}\;, \\
\text{cluster} & \quad\leftrightarrow\quad 0 \ \text{of the } \lambda_{\psi,i} \ \text{is positive}\; . 
\end{align*}
After subtracting the trace of the tidal tensor, the ellipticities are readily obtained for elliptical galaxies via Eq. (\ref{eq:linear_model}).

For spiral galaxies we have to relate the mass tensor, $-\delta_{\alpha\beta}$, to the inertial tensor by means of Eq. (\ref{eq:inertial_tensor}). In order for $I_{\alpha\beta}$ to be well defined the eigenvalues of the mass tensor need to be positive, i.e. the object under consideration has to be at a maximum of the density field. 
We then need to rotate $I_{\alpha\beta}$ from the mass tensor's eigenframe into the physical frame. The angular momentum is then obtained via Eq. (\ref{eq:ang_moment}) and the ellipticity through Eq. (\ref{eq:quadratic_model}).

For both models the correlation functions $\langle\epsilon_\pm \epsilon'_\pm\rangle (|\boldsymbol{x} - \boldsymbol{x}'|)$ are estimated and the contribution of each environment to the total alignment signal can be derived: This is achieved by computing the variance of the ellipticity field at a fixed separation $|\boldsymbol{x} - \boldsymbol{x}'|$ by generating samples of the tidal shear field, the inertia field, and consequently of the ellipticities at two locations in the large-scale structure, while differentiating between different environments by counting the number of positive eigenvalues of the tidal shear tensor. It should be noted that the relative contribution will not depend on the typical mass of a galaxy in case of the linear alignment model (\ref{eq:linear_model}), since it will only give a different overall factor $\sigma_0$. Furthermore, the evolutionary factors in front of the angular moment in Eq. (\ref{eq:ang_moment}) will drop out as well as only the normalized angular momentum enters in the ellipticity expression. Therefore, we expect no time evolution in this type of alignment either.

\begin{figure}
\begin{center}
\includegraphics[width = 0.475\textwidth]{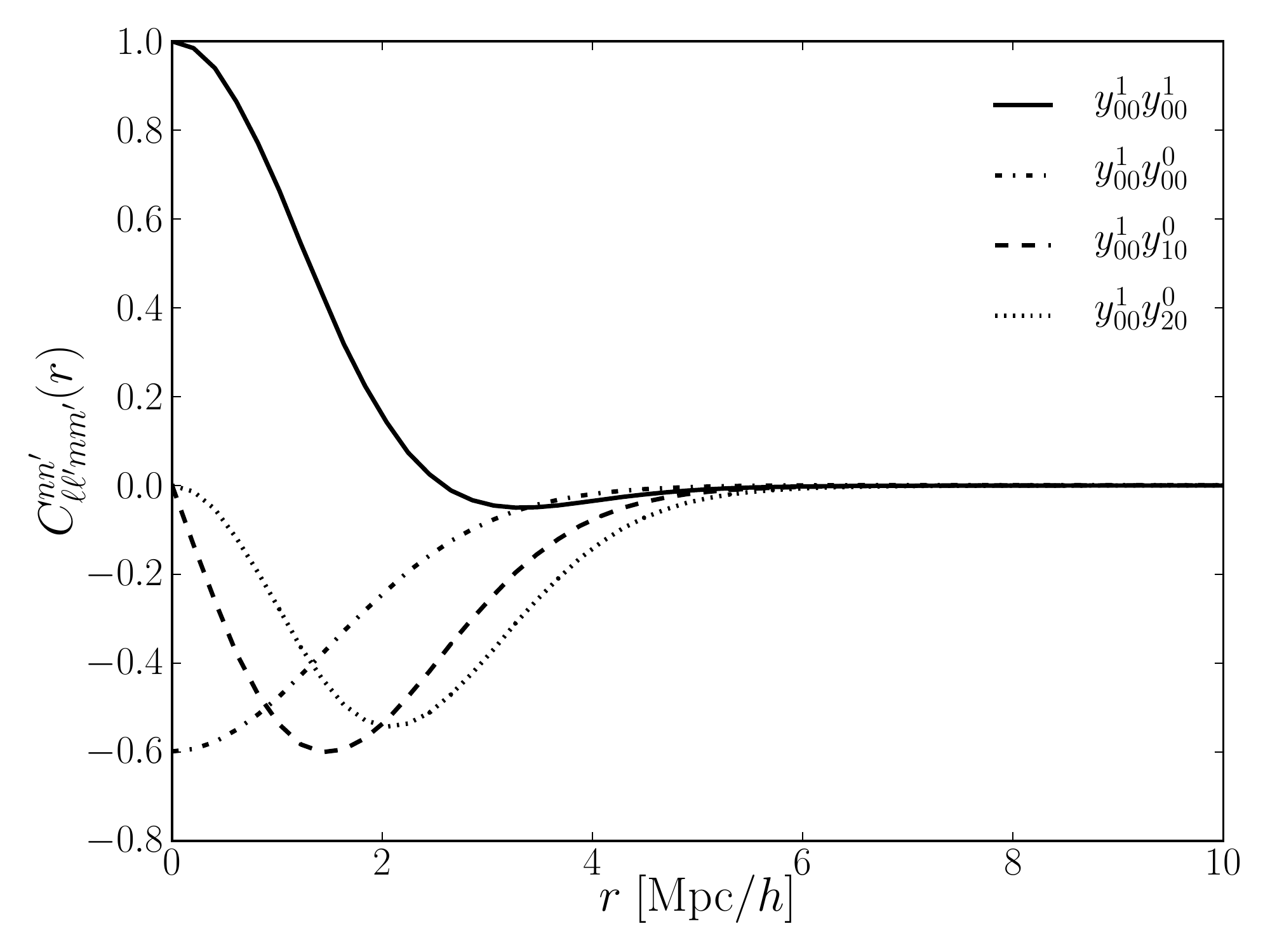}
\caption{Correlation function from Eq. (\ref{eq:correlation_matrix_components}) for different combinations of $n,\ell,m$.} 
\label{fig:ellipticals_contribution}
\end{center}
\end{figure}

\section{Results}\label{sec:results}
In this section we discuss the results obtained for the two alignment models discussed in \autoref{sec:shapecorrelations}.

\subsection{Elliptical galaxies}
The left column of \autoref{fig:spirals_contribution} shows the relative abundance of each combination of environments as a function of radial distance. This selection is purely based on the criteria introduced in \autoref{sec:sampling} with the values of the mass tensor being marginalized over. Clearly it is least likely to find a cluster and a void, especially at small distances. Most objects will be situated at a sheet or filament, which is easily understood from the number of possible combinations of eigenvalues for these environments. Furthermore we note the degeneracy between for example cluster-cluster and void-void. This effect is due to the Gaussianity of the random field, yielding positive values equally likely as negative values. The situation is therefore symmetric.

\begin{figure}
\begin{center}
\includegraphics[width = 0.95\textwidth]{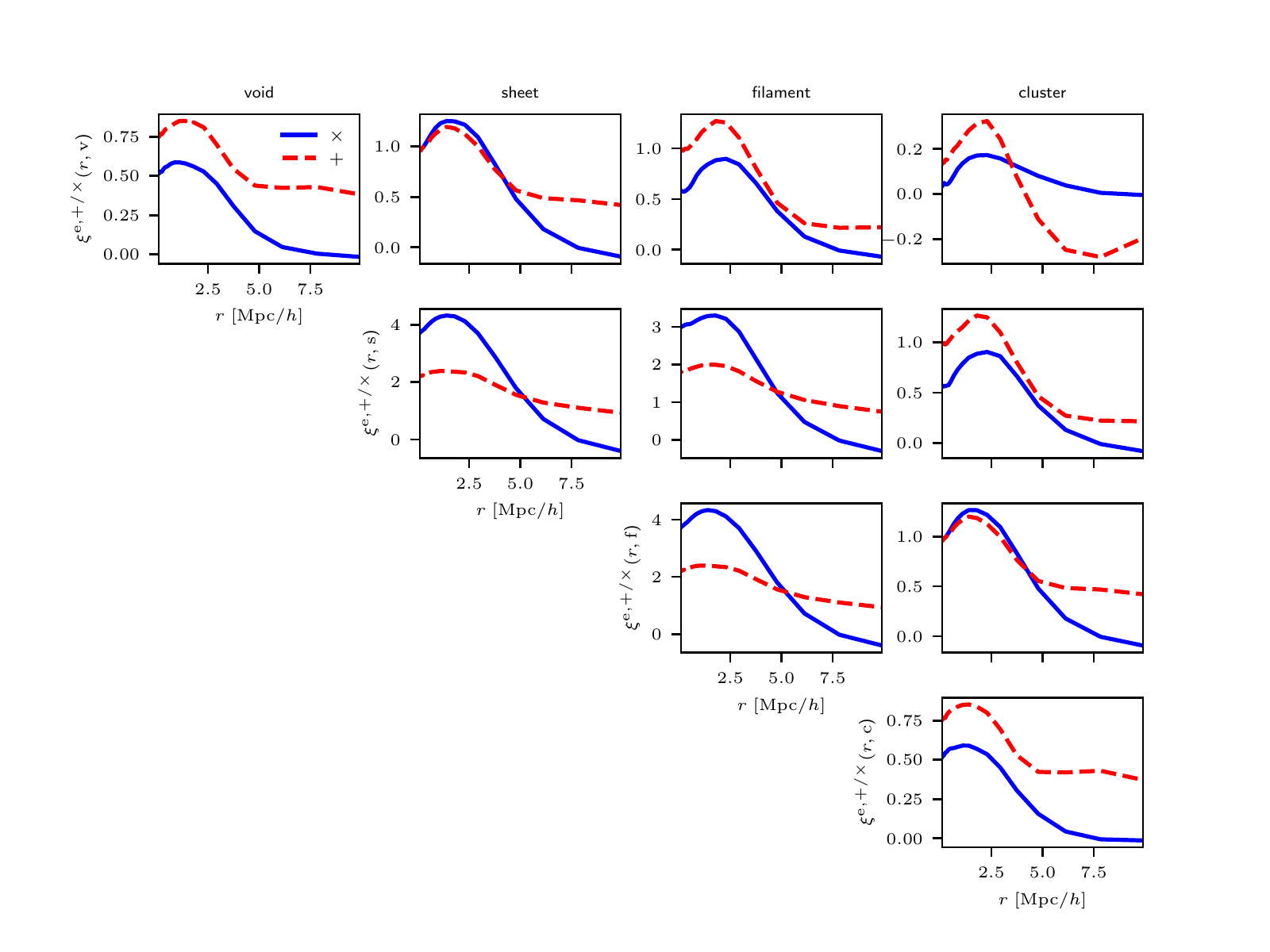}
\caption{Ellipticity correlation $\langle\epsilon_+\epsilon^\prime_+\rangle$ (dashed curve) and $\langle\epsilon_\times\epsilon^\prime_\times\rangle$ (solid curve) as a function of radius. The correlation function depends on the environment and the structure of the plot is as follows: The first row correlates elliptical galaxies situated in voids with the other types. Consequently, the upper left plot corresponds to the shape correlation induced by two galaxies both appearing in a void. In contrast the second plot in the first row shows the shape correlation between two galaxy with one located in a void and the other in a sheet. Furthermore, the environments are abbreviated by their first letter.}
\label{fig:correlation function_ellipticals}
\end{center}
\end{figure}

\autoref{fig:correlation function_ellipticals} shows the estimated correlation function in different in the different environments. The corner plot is structured as follows: on the diagonal one finds the auto-correlation function of the different environments. Off-diagonal plots show the cross-correlation between two different environments. For example the second plot in the first row shows the ellipticity correlation induced by a pairs of galaxies with one of them situated at a sheet, while the other one is located in a void. Furthermore, the correlation functions 
\begin{equation}
\xi^{\mathrm{e},+/\times}(r) \equiv\langle\epsilon_{+/\times}(\boldsymbol{x}) \epsilon_{+/\times}(\boldsymbol{x}')\rangle\;,
\end{equation}
are shown in red (dashed) and blue (solid) respectively. Here, $r = |\boldsymbol{x} - \boldsymbol{x}'|$.

We find that the strongest shape correlations is induced by sheets and filaments. This can be easily understood from the linear alignment model: Being elongated structures, sheets and filaments will lead to a stronger elongation of the object subjected to the tidal field. The amount of alignment is roughly eight times as high as for clusters and voids. Furthermore, we again find that voids and clusters behave identically, and that the same is true for filaments and sheets. It should also be noted that $\xi^+$ is larger than $\xi^\times$ for voids and clusters, while the situation is reversed for filaments and sheets. Interestingly, $\xi^+$ changes from a correlation to an anti-correlation when considering voids together with clusters. Lastly, we find that $\xi^+$ has a larger correlation length than $\xi^\times$ for all cases.

In summary, the intrinsic alignment signal of elliptical galaxies will be dominated by nearly an order of magnitude by the ellipticity correlations of objects situated at voids and filaments. This is also enhanced by the fact that sheets and filaments are more abundant than clusters or voids, in the sense that filament regions make up a large portion of the cosmic volume.

\subsection{Spiral galaxies}
\begin{figure}
\begin{center}
\includegraphics[width = 0.475\textwidth]{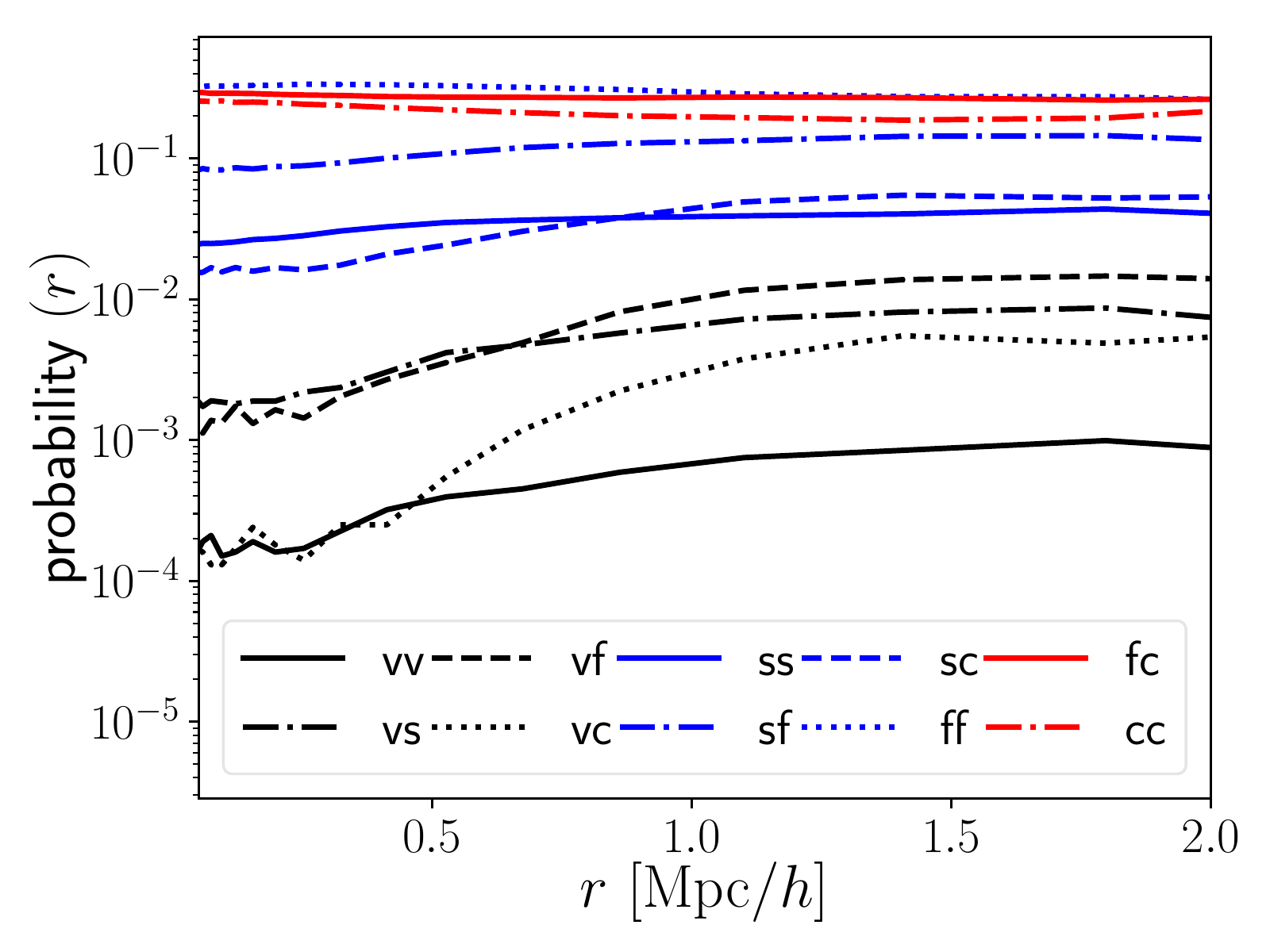}
\caption{Probability of finding a combination of two environments separated by a distance $r$ including the maximum constraint on the mass tensor.}
\end{center}
\label{fig:spirals_contribution}
\end{figure}

We repeat the same exercise as for ellipticals for spirals and show the abundance of the different environments for in the right column of \autoref{fig:spirals_contribution}.

\begin{figure}
\begin{center}
\includegraphics[width = 0.475\textwidth]{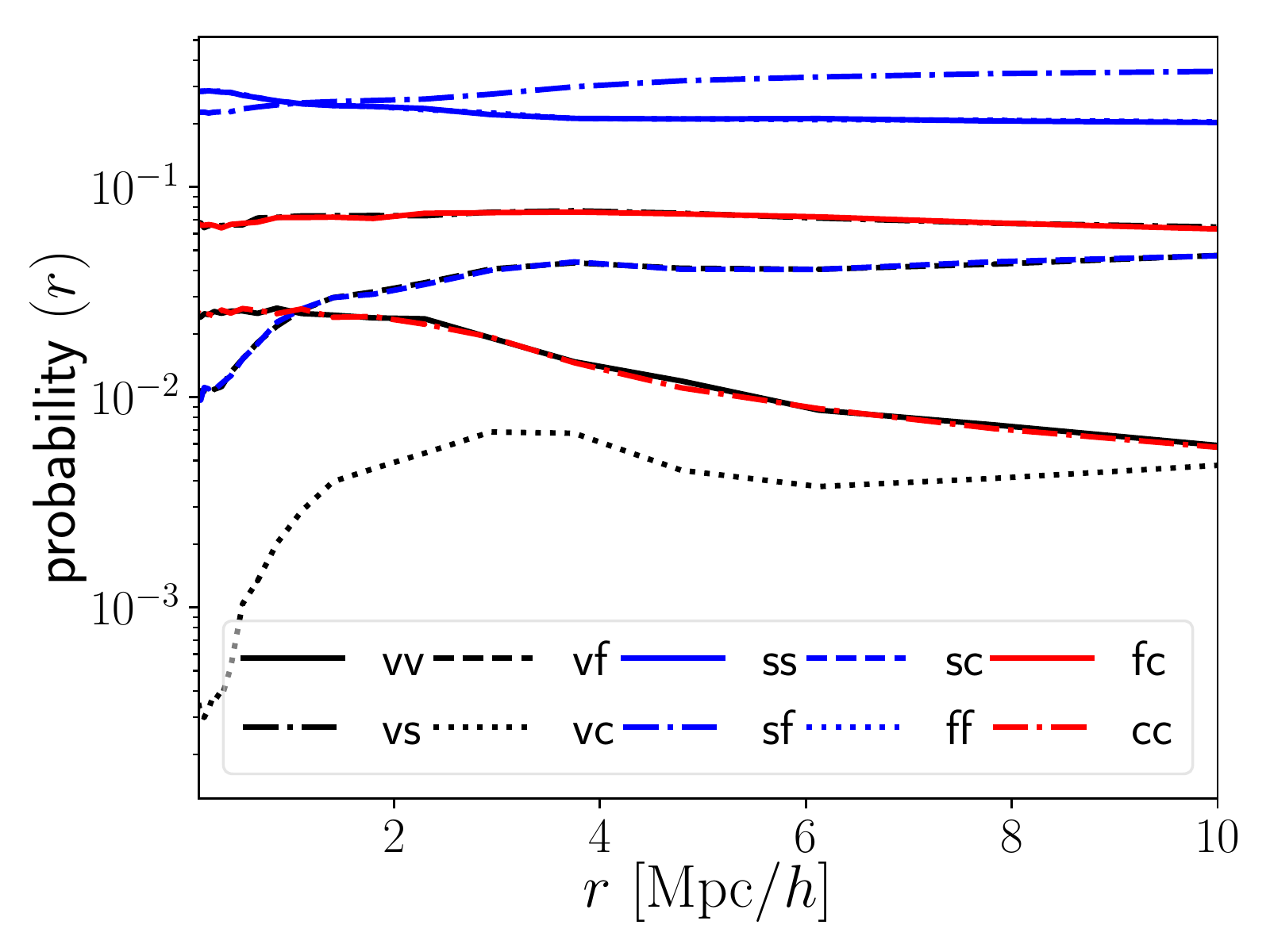}
\caption{Probability of finding a combination of two environments separated by a distance $r$ without including the maximum constraint on the mass tensor.}
\end{center}
\label{fig:ellipticals_contribution}
\end{figure}

The overall trend of the curves is the same as for the elliptical galaxies. However,the first thing to notice is the breakup of the degeneracies which were present in \autoref{fig:ellipticals_contribution}. Especially clusters are now very likely to be sampled in comparison to voids. This is because samples with negative semi-definite mass tensor are rejected which destroys the symmetry of the Gaussian distribution from which the samples are drawn. In other words, it is very unlikely to find a peak of the density field in a void. It should finally be noted that the relative abundance of each environment does not depend on the mass scale chosen, since both, peak and environment constraint, depend solely on the sign of the eigenvalues but not on their absolute magnitude.

\begin{figure}
\begin{center}
\includegraphics[width = 0.95\textwidth]{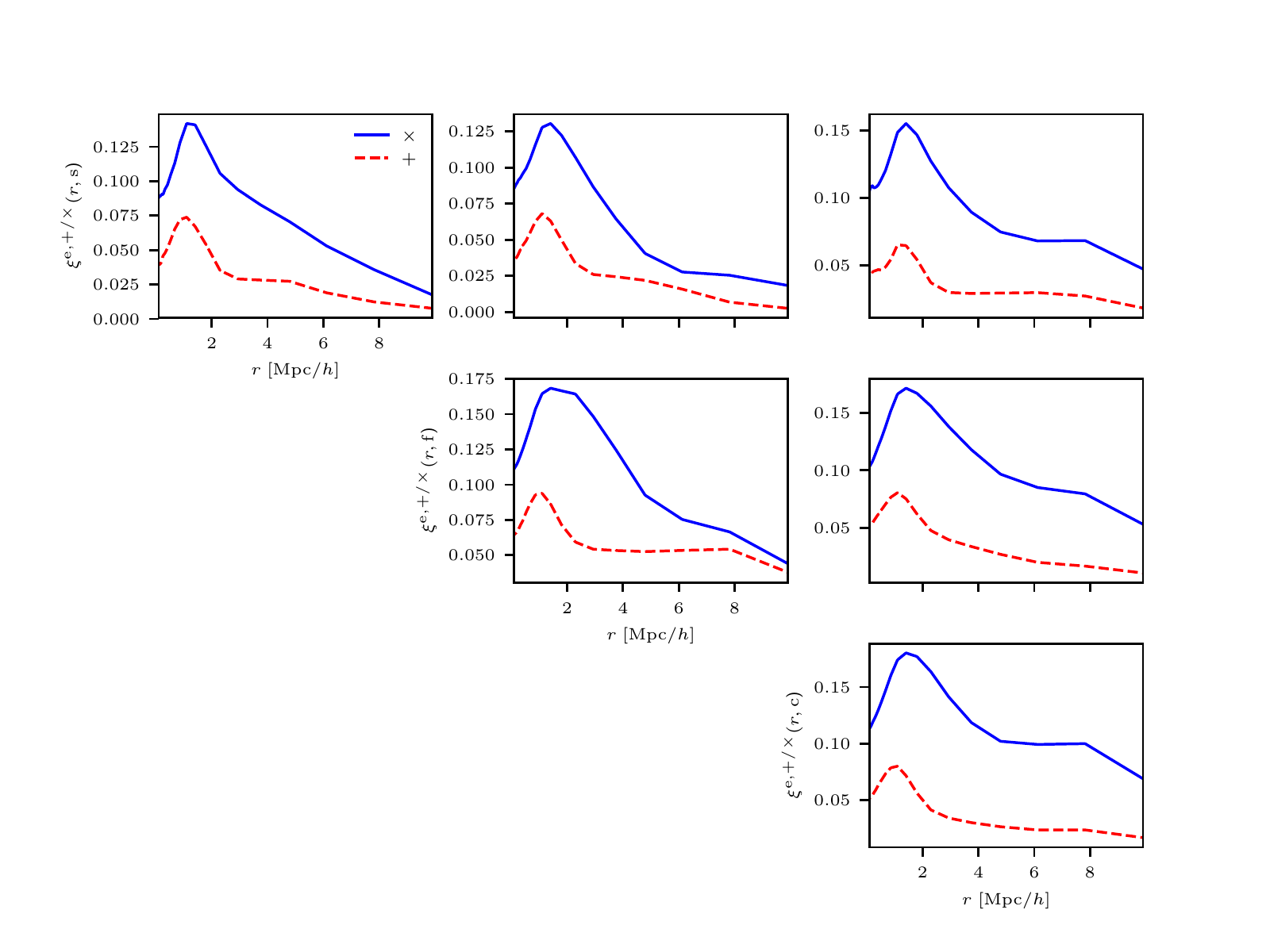}
\caption{Ellipticity correlation $\langle\epsilon_+\epsilon^\prime_+\rangle$ (dashed curve) and $\langle\epsilon_\times\epsilon^\prime_\times\rangle$ (solid curve) as a function of radius for elliptical galaxies. The structure is the same as in \autoref{fig:correlation function_ellipticals} with the first row left out. A mass scale $M = 10^{11} \mathrm{M}_\odot h^{-1}$ was used.}
\label{fig:correlation_spirals}
\end{center}
\end{figure}

In \autoref{fig:correlation_spirals} we show the correlation function for spiral galaxies. It should be noted that we ignore contributions including a void, since they are very unlikely (compare \autoref{fig:correlation_spirals}). Furthermore, we would like to stress that one should not compare the overall amplitude of the correlation functions in \autoref{fig:correlation_spirals} and \autoref{fig:correlation function_ellipticals} since we did not fix the coupling constants in the two models. Therefore, both figures can only show relative magnitudes between the different environments, but not between the two alignment models. First of all we note that $\xi^\times$ is larger than $\xi^+$ for all environments. Furthermore, we find that the correlation length of the intrinsic ellipticities for spiral galaxies is smaller than for elliptical galaxies as one would expect due to the influence of $\delta_{\alpha\beta}$ on the observed ellipticity. Additionally we find that, in contrast to elliptical galaxies, the strength of the alignment signal is comparable in all environments. This is due to the fact that the alignment in this case arises from the orientation of the galaxy only. Finally, we like to stress again, that there is no mass-dependence in this model since the ellipticity only depends on the normalised angular momentum.

\begin{figure}
\begin{center}
\includegraphics[width = 0.45\textwidth]{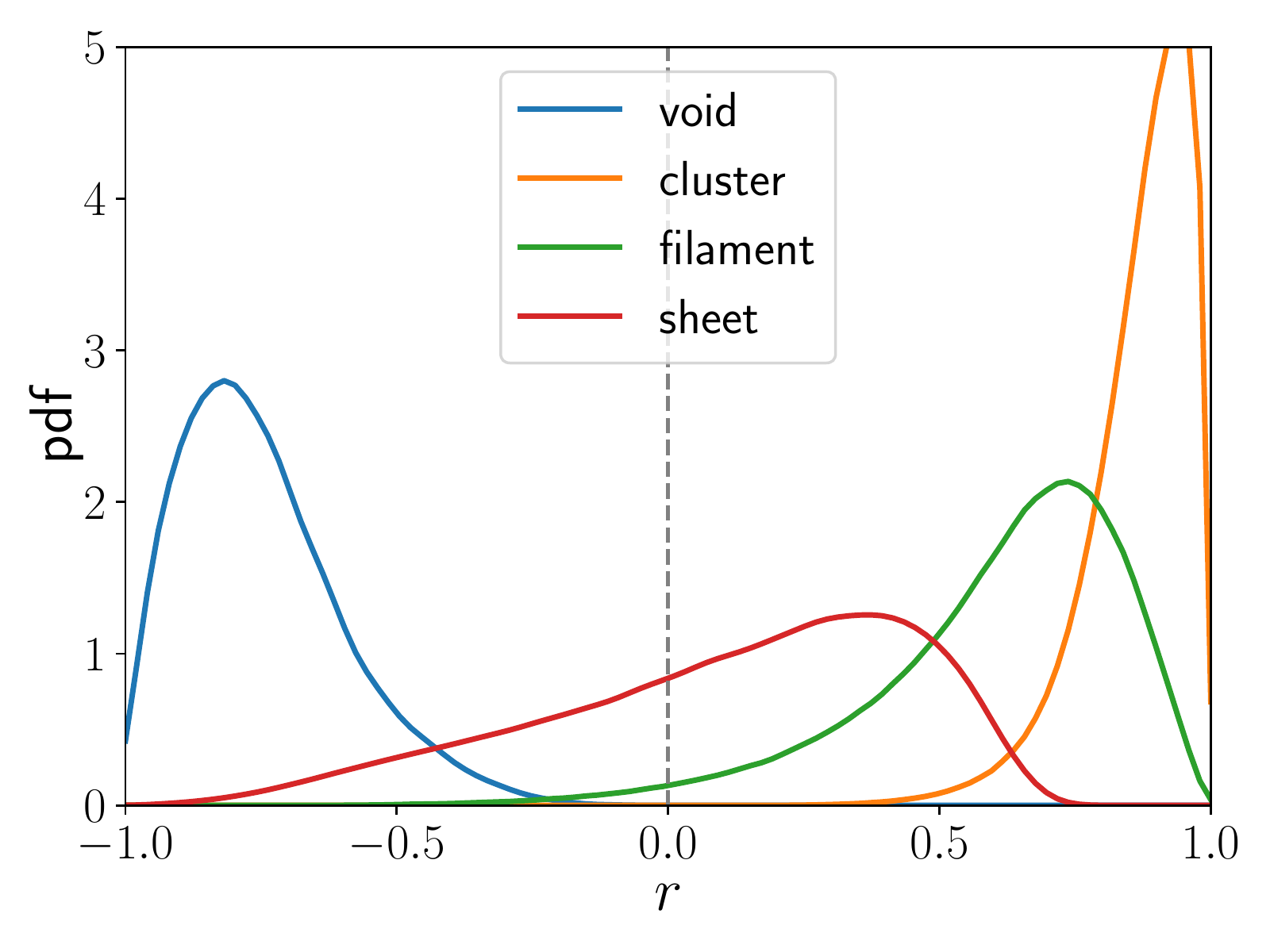}
\includegraphics[width = 0.45\textwidth]{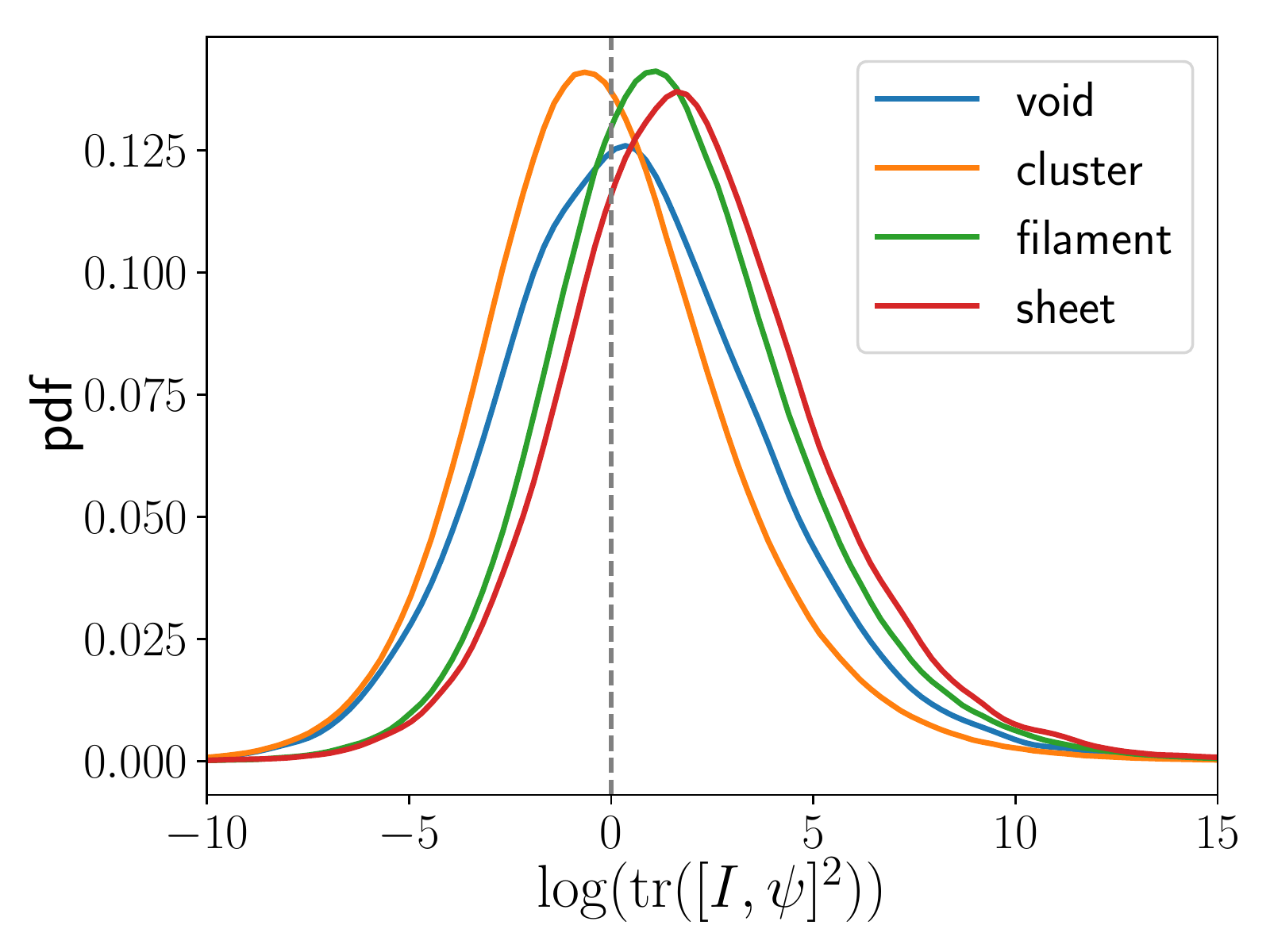}
\caption{Probability density functions for the Pearson correlation coefficient (right) and the trace of the squared commutator of the inertial tensor and the tidal tensor distinguished by environment. The employed mass scale is $M = 10^{11}M_\odot/h$.}
\label{fig:measure_misalignment}
\end{center}
\end{figure}

A central mechanism for the generation of angular momenta by tidal torquing is a misalignment of the shear and inertia eigensystems. As suggested by eqn.~\ref{eq:ang_moment}, the eigensystems of $\boldsymbol{\psi}$ and $\boldsymbol{I}$ need to be different for angular momentum to be generated. This mechanism is captured by a misalignment parameter $a$ in the model proposed by \citep{lee_cosmic_2000}, which predicts the angular momentum direction directly from the tidal shear. As a possible way of quantifying the degree of misalignment between the eigensystems of $\boldsymbol{\psi}$ and $\boldsymbol{I}$ is provided by the Cauchy-Schwarz-inequality,
\begin{equation}
\tr\left(\boldsymbol{I \psi}\right) \leq \sqrt{\tr\left(\boldsymbol{I}^2\right)\tr\left(\boldsymbol{\psi}^2\right)},
\end{equation}
such that $\tr\left(\boldsymbol{I \psi}\right)$ in units of $\sqrt{\tr\left(\boldsymbol{I}^2\right)\tr\left(\boldsymbol{\psi}^2\right)}$ is bounded from below by $-1$ and from above by $+1$. For identical eigensystems one would recover the equality of both terms. In order to quantify the efficiency of tidal torquing in a Gaussian random field in different environments, we provide samples of $\boldsymbol{I}$ and of $\boldsymbol{\psi}$ and compute the Pearson-correlation coefficient $r$,
\begin{equation}
r = \frac{\tr\left(\boldsymbol{I \psi}\right)}{\sqrt{\tr\left(\boldsymbol{I}^2\right)\tr\left(\boldsymbol{\psi}^2\right)}}
\end{equation}
differentiated by environment. The corresponding distributions of $r$ that result from the sampling process are depicted in the left panel of \autoref{fig:measure_misalignment}

Numerically, on a mass scale corresponding to $10^{11}M_\odot/h$, we obtain $\langle r \rangle \simeq -0.75$ for void regions, $\langle r\rangle \simeq 0.11$ for sheets, $\langle r\rangle\simeq 0.61$ for filaments and $\langle r\rangle \simeq 0.85$ for superclusters, indicating that tidal torquing is most effective in anisotropic regions such as filaments and sheets, because in void and cluster regions $r$ has values close to $\pm 1$, as an expression of similarity of the eigensystems of shear $\psi$ and inertia $I$.

Similarly, the degree of misalignment can be measured by computing the trace of the squared commutator, $\tr\left[\boldsymbol{I},\boldsymbol{\psi}\right]^2$, where the square is necessary as $\tr\left[\boldsymbol{I},\boldsymbol{\psi}\right] = \tr(\boldsymbol{I},\boldsymbol{\psi})-\tr(\boldsymbol{\psi},\boldsymbol{I}) = 0$. The distribution of this measure of misalignment is shown in the right panel of \autoref{fig:measure_misalignment}, as the values are sampled from a Gaussian random field smoothed on a mass scale of $10^{11}M_\odot/h$. Again, anisotropic regions like sheets and filaments show on average higher degrees of misalignment between the shear and inertia eigensystems than voids and clusters, which supports previous reports \citep{codis_spin_2015, codis_galaxy_2018} with a different derivation. 

\section{Conclusion}\label{sec:summary}
In this work we investigated the dependence of intrinsic ellipticity correlations of galaxy while differentiating the environment of the galaxies by its large-scale dynamics. We distinguished four environments: voids, sheets, filaments and clusters and examined the induced shape correlations in the linear and quadratic alignment model for elliptical and spiral galaxies, respectively. With four classes of environment it is possible to define four correlations of the ellipticity field for identical environments and six correlation functions where the two galaxies reside in different environments, adding up to ten possible correlations.

We computed ellipticity correlation functions conditional of the galaxy's location in a sampling process for a Gaussian random field, i.e. for a Gaussian probability distribution that provides properly correlated values for the tidal shear and the inertia tensor at two points in the large-scale structure. On these values, we construct samples for the ellipticity distribution and impose a differentiation on the basis of the number of positive eigenvalues of the tidal shear tensor.

We did not compare absolute values of the correlation strength, since the theoretical predictions do not fix the amplitude of the alignment, however, the relative amplitudes were compared. In fact, one would expect a constant of proportionality mediating between the shape of an elliptical galaxy and the local tidal shear field, and an effective constant that links the orientation of the angular momentum to the tidal shear field, as well as possible misalignments between the galactic disc and the angular momentum of the galaxy's host halo. Measurements of both parameters are a matter of numerical simulations and involve a physical modelling of the galaxy formation process.

Elliptical galaxies are mainly aligned in filaments and sheets. Compared to clusters and voids their intrinsic alignment signal dominated by an order of magnitude by the correlation between these two environments. Cross-correlations between different environments show a similar behaviour. The two possible correlations of the parity eigenstates of the ellipticity field show a complex behaviour in different environments, which we would explain in terms of broken isotropy of the tidal shear field, as the environment selection singles out anisotropic regions. While one observes that $\bra\epsilon_+\epsilon_+^\prime\ket$ in voids and supercluster regions dominates over $\bra\epsilon_\times\epsilon_\times^\prime\ket$, the relative magnitude of these correlations is scale dependent in sheets and filaments. In general, we encounter for the linear alignment model at a mass scale of $10^{11} M_\odot/h$ a correlation length of a few Mpc$/h$.

Spiral galaxies pick up a comparable amount of shape correlations in all types of environments and their shape correlations show similar, if not shorter correlation lengths. We are not able to sample in the Gaussian model very closeby galaxy pairs, due to imposing the smoothing scale, which is of the order of 1 Mpc$/h$, such that peak pairs at small distances are strongly suppressed. Comparing different environments one sees that spiral galaxies show similar shapes and amplitudes in their ellipticity correlation and do not depend strongly on the environment. We would argue that this is mainly due to the fact that shape correlations due to angular momentum are a pure orientation effect, where the mismatch between the eigensystems of shear and inertia matters.

Investigating the average amount of misalignment between the shear and inertia eigensystems in a more detailed way led us to computing two quantities, namely the traces $\tr(\boldsymbol{I\psi})$ and $\tr\left[\boldsymbol{I},\boldsymbol{\psi}\right]^2$. The first, $\tr(\boldsymbol{I\psi})$, is bounded by a trace inequality to be in the interval $-1\ldots+1$ if normalised by $\sqrt{\tr(\boldsymbol{I}^2)\tr(\boldsymbol{\psi}^2)}$, and the values $+1$ and $-1$ are assumed for matching eigensystems. The second, $\tr\left[\boldsymbol{I},\boldsymbol{\psi}\right]^2$ provides a scalar measure of misalignment between the eigensystems, as matching eigensystems would imply a vanishing commutator. Sampling from a Gaussian random fields shows that both measures show that larger eigensystem mismatches are found in anisotropic regions such as sheets and filaments, implying that tidal torquing is more efficient.

There is no correlation shape correlation between elliptical and spiral galaxies in Gaussian random fields: If the ellipticity of spiral galaxies follows a quadratic alignment model and that of elliptical galaxies a linear alignment model, the correlation between the two would amount to a third moment of a Gaussian deviate, which is naturally vanishing. The amplitude of ellipticity correlations where the galaxies are placed in different environments are bounded by the Cauchy-Schwarz inequality.

For both models the environmental dependence of ellipticity correlations does not depend strongly on the mass scale considered, but it would be in principle possible to impose a second mass scale on the tidal shear field generated by the environment. Both alignment models would not incorporate evolution effects of the alignment itself that are independent from the growth of structures, and hence of the external tidal shear field that is acting on the galaxy. In fact, as a consequence of the Poisson-equation, the tidal shear field scales in cosmic structure formation as $D_+(a)/a$, which is close to 1 for most of cosmic history and only deviates from 1 at low redshifts, where dark energy starts to dominate. There are, however, evolution effects of the alignment signal from peculiar motion, which forces the galaxy to experience varying tidal shear fields as it is advected by the cosmic velocity field \citep{giahi_evolution_2013, schmitz_time_2018}.

These findings can be used to further test intrinsic alignment models against numerical simulations and in actual data. Furthermore, it could be helpful to distinguish between effects induced by intrinsic alignments and gravitational lensing, as the latter does not depend on the environment around the sources with the caveat that studies of the environment are notoriously difficult to carry out observationally. Finally, an interesting question is the interplay between intrinsic alignment and weak gravitational lensing in the environmental dependence of the GI-alignment terms: At least in this cross correlation, lensing would be agnostic about environment, and elliptical galaxies would show a preference of high values for ellipticity in sheets and filaments.

\bibliographystyle{JHEP}
\bibliography{library}

\acknowledgments
RR acknowledges financial support of the Helmholtz-foundation and of the HEiKA research partnership between Heidelberg University and the KIT in Karlsruhe. We thank Robert Lilow for helpful discussions.

\end{document}